\documentclass[12pt]{article}
\usepackage{amssymb}
\usepackage{epsfig}
\usepackage{float}
\usepackage{graphicx}
\usepackage{amsmath}
\newcommand{\nua}[1]{\ensuremath{\rlap
           {\kern-2.5pt\ensuremath
           {\overset{\scriptscriptstyle(-)}{\phantom{\nu}}}}
           {\ensuremath{{\nu}_{#1}}}}}

\begin{document}

\begin{center}
{\bf Neutrino and The Standard Model}
\end{center}
\begin{center}
S. M. Bilenky
\end{center}

\begin{center}

{\em  Joint Institute for Nuclear
Research, Dubna, R-141980, Russia;\\}
{\em TRIUMF 4004 Wesbrook Mall, Vancouver BC, V6T 2A3 Canada\\}
\end{center}
\begin{abstract}
After discovery of the Higgs boson at CERN  the Standard Model acquired a status of the full, correct theory of the elementary particles in the electroweak range. What general conclusions can be inferred from the SM? I am suggesting here that in the framework of such general principles as local gauge symmetry, unification of the weak and electromagnetic interactions and Brout-Englert-Higgs  spontaneous breaking of the electroweak symmetry  nature chooses
the simplest possibilities. It is very plausible that massless  left-handed  neutrinos (simplest, most economical possibility) play crucial role in the determination of the charged current structure of the Standard Model and that neutrino properties (masses and nature) are determined by a beyond the Standard Model physics. The discovery of the neutrinoless double $\beta$-decay and  proof that neutrinos with definite masses are Majorana particles would be an important evidence in favor  of the considered scenario.
\end{abstract}

\section{Introduction}
The Standard Model  is one of the greatest achievement of the physics of the XX's century. It emerged as a result of numerous experiments and such fundamental theoretical principles as local gauge invariance, unification of the weak and electromagnetic interactions, spontaneous breaking of the electroweak symmetry.

 After the discovery of the Higgs boson at LHC the Standard Model was established as full, correct theory of physical phenomena in the electroweak energy scale (up to about 300 GeV). We will try here to discuss some lessons which we can be inferred from the Standard Model

 There are many questions connected with the Standard Model: why left-handed and right-handed quark, lepton and neutrino fields have different transformation properties, why in unified electroweak interaction the weak part maximally violate parity and the electromagnetic part conserve parity,
 why mixing takes place etc. I suggest here that {\em the charged current structure of the SM is determined by neutrinos.}

 {\em From my point of view the SM started with the theory of the massless two-component neutrino.} In 1929 soon after Dirac proposed his famous equation,  Weyl \cite{Weyl}   introduced two-component spinors $\psi_{L,R}(x)$ as a simplest and most elegant possibility for massless particles. The two-component spinors  are determined by the relations $\gamma _{5}\psi_{L,R}(x)=\mp \psi_{L,R}(x)$ and satisfy the  equations
\begin{equation}\label{weyl}
i\gamma^{\alpha}\partial_{\alpha}~\psi_{L,R}(x)=0.
\end{equation}
However, the two-component fields for massless particles were rejected at that time because the equations (\ref{weyl}) do not conserve parity.\footnote{When Weyl introduced two-component spinors and discussed them in some details in \cite{Weyl} he, probably, followed his principle: "My work  always tried to unite the truth with the beautiful, but when I had to choose one or the other, I usually chose the beautiful."} Pauli in his book on Quantum Mechanics \cite{Pauli} wrote "...because the equation for
$\psi_{L}(x)$ ($\psi_{R}(x))$ is not invariant under space reflection it is not applicable to the physical reality".

After  in 1957 large violation of parity in the $\beta$-decay and other weak processes  was discovered \cite{Wu,Lederman} Landau \cite{Landau}, Lee and Yang \cite{LeeYang} and Salam \cite{Salam} from different reasons proposed two-component theory for neutrino. According to this theory neutrino mass is equal to zero and neutrino field $\nu_{L}(x)$ (or
$\nu_{R}(x)$)  satisfies the equation
\begin{equation}\label{weyl1}
i\gamma^{\alpha}\partial_{\alpha}~\nu_{L}(x)=0 ~~~(\mathrm{or}~~~i\gamma^{\alpha}\partial_{\alpha}~\nu_{R}(x)=0).
\end{equation}
If neutrino is the two-component massless $\nu_{L}$ (or $\nu_{R}$) Weyl particle in this case

\begin{enumerate}
  \item Large violation of the parity in the $\beta$-decay and other weak processes must be observed (in agreement with the results of the  Wu et al and other experiments \cite{Wu,Lederman}).

  \item Neutrino   helicity is equal -1 (+1) and antineutrino helicity is equal to +1 (-1).
\end{enumerate}
The point 1. is obvious. In fact, under the space inversion the left-handed (right-handed) neutrino field is transformed into the right-handed (left-handed) field
\begin{equation}\label{inversion}
\nu'_{R,L}(x')=\eta\gamma^{0}\nu_{L,R}(x).
\end{equation}
Here $x'=(x^{0},-\vec{x})$ and $\eta$ is a phase factor. Thus, if neutrino field is $\nu_{L}(x)$ (or $\nu_{R}(x)$) the Lagrangian is not invariant under the space inversion.

In order to see that neutrino is a particle with definite helicity  let us consider the spinor $u^{r}(p)$ which describes a massless particle with the momentum $p$ and helicity $r$. We have
\begin{equation}\label{helicity}
\vec{\Sigma}\cdot \vec{n}~u^{r}(p)=r~u^{r}(p),\quad \gamma\cdot p ~ u^{r}(p)=0.
\end{equation}
Here $\vec{\Sigma}=\gamma_{5}\gamma^{0}\vec{\gamma}$  is the operator of the spin and $\vec{n}=\frac{\vec{p}}{|\vec{p}|}$ is the unit vector in the direction of the momentum. From (\ref{helicity}) it follows that
\begin{equation}\label{helicity1}
 \gamma_{5}~u^{r}(p)=r~u^{r}(p)~~\mathrm{and}~~ \frac{1}{2}(1\mp \gamma_{5})~u^{r}(p)=\frac{1}{2}(1\mp r)~u^{r}(p)
\end{equation}
 Thus, if neutrino field is $\nu_{L}(x)$ ($\nu_{R}(x)$) in this case $r=-1~ (r=+1)$. Analogously, it is easy to show that antineutrino helicity is equal to +1 (-1) in the case if neutrino field is
$\nu_{L}(x)$ ($\nu_{R}(x)$).

 The neutrino helicity was measured in the spectacular
Goldhaber et al experiment \cite{Goldhaber}.  In this experiment the neutrino helicity was obtained from the measurement of the circular polarization of $\gamma$'s produced in the chain of reactions
\begin{eqnarray}\label{helicity2}
e^- + ^{152}\rm {Eu} \to \nu + \null & ^{152}\rm {Sm}^* & \nonumber
\\
& \downarrow & \nonumber
\\
& ^{152}\rm{Sm} & \null + \gamma.\nonumber
\end{eqnarray}
The authors of the paper  \cite{Goldhaber} concluded : "... our result is compatible with 100\% negative helicity  of neutrino emitted in orbital electron capture".

{\em Thus, the Goldhaber et al experiment confirmed the two-component neutrino theory.} It was shown that from two possibilities ($\nu_{L}(x)$ or $\nu_{R}(x)$) nature choose  the first one.

The number of degrees of freedom of the Weyl field is two times smaller than the number of the degrees of freedom of the four-component Dirac field. After the discovery of the large violation of parity in the $\beta$-decay and other weak processes and the measurement of the neutrino helicity it became very plausible that  {\em for the neutrino nature choose this simplest possibility.}

Let us notice that at the time when  the two-component neutrino theory was proposed it was unknown that exist three  types of neutrino. In 1962 in the Brookhaven experiment \cite{Brookhaven} it was shown that muon and electron neutrinos $\nu_{e}$ and $\nu_{\mu}$ are different particles. In 2000 the third neutrino $\nu_{\tau}$ was discovered in the DONUT experiment \cite{Donut}.

\section{On the Standard Model}

{\em The Standard Model is based on the local gauge symmetry} which is a natural symmetry for the Quantum Field Theory with depending on $x$  quantum fields.  In order to include charged leptons and quarks  the symmetry group must be non-Abelian.

The simplest possibility is $SU_{L}(2)$ group with three lepton doublets
\begin{eqnarray}
\psi^{lep}_{eL}=\left(
\begin{array}{c}
\nu'_{eL} \\
e'_L \\
\end{array}
\right),~\psi^{lep}_{\mu L}=\left(
\begin{array}{c}
\nu'_{\mu L} \\
\mu'_L \\
\end{array}
\right),~\psi^{lep}_{\tau
 L}=\left(
\begin{array}{c}
\nu'_{\tau L} \\
\tau'_L \\
\end{array}
\right),\label{lepSU2dub}
\end{eqnarray}
and three quark doublets
\begin{eqnarray}
\psi_{1L}=\left(
\begin{array}{c}
u'_{L} \\
d'_L\\
\end{array}
\right),~ \psi_{2 L}=\left(
\begin{array}{c}
c'_{L}\\
s'_L \\
\end{array}
\right),~ \psi_{3 L}=\left(
\begin{array}{c}
t'_L\\
b'_L\\
\end{array}
\right)\label{quarkSU2dub}
\end{eqnarray}
Here the lepton fields ($e'_{L}(x),\mu'_{L}(x),\tau'_{L}(x)$) and fields of the quarks with the charges $(2/3)$ and
$(-1/3)$ ($u'_{L}(x), c'_{L}(x), t'_{L}(x)$ and $d'_{L}(x), s'_{L}(x), b'_{L}(x)$), {\em like neutrino fields}, are massless two-component Weyl fields.

In order to insure the invariance under the local gauge transformations
\begin{equation}\label{locSU2}
(\psi^{lep}_{l})'(x) = e^{i\,\frac{1}{2}~\vec\tau\cdot\vec\Lambda(x)} ~\psi^{lep}_{l}(x)~ (l=e,\mu,\tau)\quad
\psi_{a}'(x) = e^{i\,\frac{1}{2}~\vec\tau\cdot\vec\Lambda(x)} ~\psi_{a}(x)~(a=1,2,3)
\end{equation}
($\Lambda_{i}(x)$ (i=1,2,3) are {\em arbitrary  functions of $x$}) we need to assume that neutrino-lepton and quark fields interact with massless vector (isovector) field $\vec A_{\alpha}(x)$ and in the free Lagrangian derivatives of the fermion fields are changed by the covariant derivatives
\begin{equation}\label{1covder}
\partial_{\alpha} \,\psi^{lep}_{lL}(x)\to  (\partial_{\alpha} +
i\,g\,\frac{1}{2}\,\vec\tau\cdot\vec A_{\alpha}(x))\,\psi^{lep}_{lL}(x),~ \partial_{\alpha} \,\psi_{aL}(x)\to  (\partial_{\alpha} +
i\,g\,\frac{1}{2}\,\vec\tau\cdot\vec A_{\alpha}(x))\,\psi_{aL}(x),
\end{equation}
where $g$ is a dimensionless interaction constant. Because the interaction constant $g$ enters into the strength tensor of the field $\vec A_{\alpha}(x)$
\begin{equation}\label{stresstenz}
\vec F_{\alpha \beta}(x)= \partial_{\alpha}\,\vec A_{\beta}(x)
-\partial_{\beta}\,\vec A_{\alpha}(x)-g\,\vec
A_{\alpha}(x)\times\vec A_{\beta}(x).
\end{equation}
{\em it must be the same for all doublets}. As a result, we come to the following  Lagrangian of the universal interaction
of fundamental fermions and vector bosons
\begin{equation}\label{interL}
\mathcal{L}_{I}(x) = -g~\vec{j}_{\alpha}\vec{A}^{\alpha},
\end{equation}
where
\begin{equation}\label{isocurrent}
\vec{j}_{\alpha}=\sum_{l=e,\mu,\tau}\bar\psi^{lep}_{lL}\gamma_{\alpha}\frac{1}{2}\vec{\tau} \psi^{lep}_{lL}+\sum^{3}_{a=1}\bar\psi_{aL}\gamma_{\alpha}\frac{1}{2}\vec{\tau} \psi_{aL}
\end{equation}
is the isovector current of lepton and quarks.

The expression (\ref{interL}) can be written in the form
\begin{equation}\label{interL1}
\mathcal{L}_{I}(x) = \left( -\frac{g}{2\,\sqrt{2}}\,
j^{CC}_{\alpha}(x)\,W^{\alpha}(x) + \rm{h.c}\right)
-g\,j^{3}_{\alpha}(x)\,A^{3\alpha }(x)~.
\end{equation}
Here
\begin{equation}\label{3CCcurrent}
j^{CC}_{\alpha}=2(j^{1}_{\alpha}+ij^{2}_{\alpha}) = 2\sum_{l=e,\mu,\tau}\bar\nu'_{lL}\gamma_{\alpha} l'_L+2\sum_{q_{1}=u,..q_{2}=d,..}\bar q'_{1L}\gamma_{\alpha} q'_{2L}
\end{equation}
is the charged current of the  leptons and quarks  and $W_{\alpha}=\frac{A^{1}_{\alpha}+iA^{2}_{\alpha}}{\sqrt{2}}$ is the field of charged, vector $W^{\pm}$ bosons.

Let us stress the following
\begin{enumerate}
  \item The interaction (\ref{interL}) is the simplest possibility. {\em It is the minimal interaction compatible with two-component neutrino theory and  local gauge invariance.}
  \item Because the interaction constant $g$ is the same for all doublets, the CC weak interaction is $(\nu_{e}, e)- (\nu_{\mu}, \mu)-(\nu_{\tau}, \tau)$ -universal.
  \end{enumerate}
{\em The Standard Model is the unified theory of the weak and electromagnetic interactions}. In the electromagnetic current the left-handed and right-handed fields enter. For example, the electromagnetic lepton current
has the form
\begin{equation}\label{EMcur}
j^{\mathrm{EM}} _{\alpha}= \sum_{l}(-1)~\bar l'\gamma_{\alpha} l'=\sum_{l}(-1)~\bar l_{L}'\gamma_{\alpha} l'_{L}
+\sum_{l}(-1)~\bar l_{R}'\gamma_{\alpha} l'_{R}.
\end{equation}
Thus, in order to unify the weak and electromagnetic interactions we must enlarge the symmetry group. A new symmetry group must include not only transformations of left-handed fields but also transformations of right-handed fields {\em of charged leptons and quarks.} At this point there is a fundamental difference between neutrinos and other  fermions: neutrinos charges are equal to zero, there is no electromagnetic current of neutrinos. The   unification of the weak and electromagnetic interactions does not  require right-handed neutrino fields. Thus, {\em a minimal possibility is to assume that  there are no right-handed neutrino fields in the SM.}

 The minimal enlargement of the $SU_{L}(2)$ symmetry group is the direct product $SU_{L}(2)\times U_{Y}(1)$ where $U_{Y}(1)$ is the Abelian group of the hypercharge.
In order to ensure local gauge $SU_{L}(2)\times U_{Y}(1)$ invariance we need to change in the free Lagrangian derivatives of left-handed and right-handed fields by the covariant derivatives
\begin{eqnarray}\label{covderiv}
\partial_{\alpha}\psi^{lep}_{lL} \to (\partial_{\alpha} +ig\frac{1}{2}\vec{\tau}\cdot\vec{A_{\alpha}}+ig'\frac{1}{2}Y^{lep}_{L}B_{\alpha})\psi^{lep}_{lL},\nonumber\\
\partial_{\alpha}\psi_{aL} \to (\partial_{\alpha} +ig\frac{1}{2}\vec{\tau}\cdot\vec{A_{\alpha}}+ig'\frac{1}{2}Y_{L}B_{\alpha})\psi_{aL}
\end{eqnarray}
and
\begin{eqnarray}\label{covderiv1}
&&\partial_{\alpha}l'_{R} \to (\partial_{\alpha} +ig'\frac{1}{2}Y^{lep}_{R}B_{\alpha})l'_{R}\nonumber\\
&&\partial_{\alpha}q'_{R} \to (\partial_{\alpha} +ig'\frac{1}{2}Y^{up}_{R}B_{\alpha})q'_{R},~~ q'_{R}=u'_{R}, c'_{R}, t'_{R}\nonumber\\
&&\partial_{\alpha}q'_{R} \to (\partial_{\alpha} +ig'\frac{1}{2}Y^{down}_{R}B_{\alpha})q'_{R},~~ q'_{R}=d'_{R}, s'_{R}, b'_{R},
\end{eqnarray}
where  $B_{\alpha}$ is the field of the vector neutral gauge bosons of the $U_{Y}(1)$ group.

 There are no constraints on the interaction constants of the Abelian $U_{Y}(1)$ local group. However, in order to unify the weak and electromagnetic interactions we need to assume that the interaction constants, correspondingly, for lepton doublets, quarks doublets, right-handed lepton singlets, right-handed up-quark singlets
and  right-handed down-quark singlets have the form
 \begin{equation}\label{interconst}
 g'\frac{1}{2}Y^{lep}_{L}, ~~~g'\frac{1}{2}Y_{L},~~~ g'\frac{1}{2}Y^{lep}_{R},~~~g'\frac{1}{2}Y^{up}_{R},~~~ g'\frac{1}{2}Y^{down}_{R}
 \end{equation}
  Here $g'$ is a constant and hypercharges of left-handed and right-handed fields of leptons and quarks $Y^{lep}_{L},Y_{L},...$ are chosen in accordance with the Gell-Mann-Nishijima relation
\begin{equation}\label{3GM-N}
    Q=T_{3}+\frac{1}{2}Y,
\end{equation}
where $Q$ is the electric charge  and $T_{3}$ is the third projection of the isotopic spin.

For the Lagrangian of the minimal interaction of lepton and quark fields with the fields of  neutral vector fields $A^{3}_{\alpha}$ and $B_{\alpha}$ we obtain the following expression
\begin{equation}\label{interL5}
\mathcal{L}^{0}_{I}=-g\,j^{3}_{\alpha}\,A^{3\alpha }-g'~\frac{1}{2}~j^{Y} _{\alpha}~B^{\alpha}.
\end{equation}
Here
 \begin{equation}\label{interL6}
 \frac{1}{2}~j^{Y} _{\alpha}=j^{EM}_{\alpha}-j^{3}_{\alpha},
 \end{equation}
where $j^{EM}_{\alpha}$ is the electromagnetic current of the leptons and quarks.

Let us notice that this last relation is due to the fact that electric charges of the left-handed components (coming from doublets) and the right-handed components (coming from singlets) are the same. Thus, if we  choose  $U_{Y}(1)$ coupling constants
in accordance with the Gell-Mann-Nishijima relation we can combine electromagnetic interaction which conserve parity with the weak interaction which violate parity into one electroweak interaction. {\em Nonconservation of parity is deeply connected with the two-component massless neutrinos.}

In order to identify in (\ref{interL5}) the Lagrangian of electromagnetic
 interaction of leptons and quarks with the electromagnetic field
\begin{itemize}
  \item instead of the fields $A^{3 \alpha}$ and $B^{\alpha}$  we need to introduce the following orthogonal "mixed" fields
 \begin{equation}\label{mixfields}
Z^{ \alpha}=\cos\theta_{W} A^{3 \alpha}-\sin\theta_{W} B^{ \alpha},\quad A^{ \alpha}=\sin\theta_{W} A^{3 \alpha}+
\cos\theta_{W}B^{ \alpha},
\end{equation}
where the  weak (Weinberg) angle $\theta_{W}$ is determined by the relation
\begin{equation}\label{Wangle}
\frac{g'}{g}=\tan \theta_{W}.
\end{equation}
\item
we need to assume  that
\begin{equation}\label{unif}
    g~\sin \theta_{W}=e,
 \end{equation}
were e is the proton charge. This relation is called the  {\em unification condition}.
\end{itemize}
Finally, the interaction Lagrangian consists of three parts:
\begin{enumerate}
  \item The Lagrangian of the interaction of the fermions with the vector field of the charged $W^{\pm}$ bosons
 \begin{equation}\label{CCinter}
\mathcal{L}^{CC}_{I} = \left( -\frac{g}{2\,\sqrt{2}}\,
j^{CC}_{\alpha}\,W^{\alpha} + \rm{h.c}\right).
\end{equation}
 \item
The Lagrangian of the interaction the fermions with the electromagnetic field  $A^{\alpha}$
\begin{equation}\label{EMinter}
\mathcal{L}^{EM}_{I} =
-j^{EM}_{\alpha}\,A^{\alpha}.
\end{equation}
  \item
The Lagrangian of the neutral current (NC) interaction the fermions with the vector field of the neutral $Z^{0}$ bosons
\begin{equation}\label{NCinter}
\mathcal{L}^{NC}_{I}=-\frac{g}{2\cos\theta_{W}}\,j^{\rm{NC}}_{\alpha}\,Z^ {\alpha},
\end{equation}
where the neutral current $j^{\rm{NC}}_{\alpha}$ is given by the expression
\begin{equation}\label{NC}
j^{\rm{NC}}_{\alpha}=2~ j^{3}_{\alpha} -2~\sin^{2}\theta_{W}\,j^{\rm{EM}}_{\alpha}.
\end{equation}
\end{enumerate}
The structure of the CC term is originated from the theory of the two-component neutrino. The structure of the NC term is determined by the unification the CC weak interaction and EM interaction on the basis of the $SU_{L}(2)\times U_{Y}(1)$ group. It is evident that the  Lagrangian of interaction of the fundamental fermions and gauge vector bosons
\begin{equation}\label{totLag}
\mathcal{L}_{I}=\mathcal{L}^{CC}_{I}+\mathcal{L}^{EM}_{I}+\mathcal{L}^{NC}_{I}
\end{equation}
{\em is the minimal, simplest Lagrangian} (in the framework of the local gauge invariance and unification).

Up to now fields of all fundamental fermions and gauge vector bosons were massless. In order to build a realistic theory of the electroweak interaction we need to use a mechanism of the generation of masses of $W^{\pm}$ and $Z^{0}$ bosons, quarks and charged leptons. The photon must remain massless. Neutrino masses is a special subject. We will discuss it later.

{\em The Standard model mechanism  of the mass generation is the Brout-Englert-Higgs mechanism} \cite{Brout,Higgs}. It is based on the phenomenon of the spontaneous symmetry breaking. The spontaneous symmetry breaking takes place in several many-body phenomena such as ferromagnetism and others. It happens if the Hamiltonian of the system has some symmetry, vacuum states are degenerated  and (breaking the symmetry) the system spontaneously occupies  one vacuum state. It was suggested  \cite{Nambu,Goldstone} that the phenomenon of the spontaneous symmetry breaking takes place also in the Quantum Field Theory.

In order to ensure the spontaneous symmetry breaking we need to assume that  {\em in addition to the fields of fundamental fermions and gauge vector bosons the scalar Higgs field  is included in the system.}

We will assume that the Higgs field
\begin{equation}\label{Hdoub}
\phi(x) ={\phi _{+}(x)\choose\phi _{0}(x)}
\end{equation}
is transformed as the $SU_{L}(2)$ doublet. Here $\phi _{+}(x)$ and $\phi _{0}(x) $ are  complex  charged and neutral scalar fields. According to  the Gell-Mann-Nishijima relation the hypercharge of $\phi(x)$ is equal to one.
We will see later that  this assumption give us the most economical, simplest (in the framework of the spontaneous breaking of the symmetry) possibility to provide masses of $W^{\pm}$ and $Z^{0}$ vector bosons.

The  part of the $SU_{L}(2)\times U_{Y}(1)$ invariant Lagrangian, in which the Higgs field enters,  has the form
\begin{equation}\label{HiggsL}
 \mathcal{L}=
((\partial_{\alpha}+i\,g\,\frac{1}{2}\,\vec\tau\cdot\vec
A_{\alpha} + i\,g'\,\frac{1}{2}\, B_{\alpha} )\,\phi
)^{\dagger}
(\partial^{\alpha}+i\,g\,\frac{1}{2}\,\vec\tau\cdot\vec
A^{\alpha} + i\,g'\,\frac{1}{2}\,
B^{\alpha} \,)\,\phi
-V(\phi^{\dagger}\,\phi).
\end{equation}
Here the term $V(\phi^{\dagger}\,\phi)$ (which is called potential) is given by the expression
\begin{equation}\label{Higgspot}
V(\phi^{\dagger}\,\phi)= -\mu^{2}\,\phi^{\dagger}\,\phi
+\lambda\,(\phi^{\dagger}\,\phi)^{2},
\end{equation}
where $\mu^{2}$ and $\lambda$ are positive constants. The constant $\mu$ has dimension of $M$ and $\lambda$ is dimensionless constant.

Existence of the Higgs field fundamentally change the property of the system: the energy of the system reaches the minimum {\em at nonzero values of the Higgs field}. In fact, the  energy of the system reaches the minimum at such values of Higgs field which minimize the potential $V(\phi^{\dagger}\,\phi)$. We can rewrite the potential $V(\phi^{\dagger}\,\phi)$ in the form
\begin{equation}\label{Higgspot1}
V(\phi^{\dagger}\,\phi)=\lambda~\left(\phi^{\dagger}\,\phi-
\frac{\mu^{2}}{2\lambda}\right)^{2}-\frac{\mu^{4}}{4\lambda}.
\end{equation}
From this expression it is obvious that the potential reaches minimum at
\begin{equation}\label{Hmin}
(\phi^{\dagger}\,\phi)_{0}= \frac{v^{2}}{2}
\end{equation}
where
\begin{equation}\label{Hmin1}
v^{2}=\frac{\mu^{2}}{\lambda}.
\end{equation}
Taking into account the conservation of the electric charge, for the vacuum values of the Higgs field we have
\begin{equation}\label{vev5}
    \phi_{0} = {0\choose\frac {v}{\sqrt{2}} }~e^{i\alpha},
\end{equation}
where $\alpha$ is an arbitrary phase. This freedom is obviously connected with the gauge symmetry of the Lagrangian.
We can choose
\begin{equation}\label{vev6}
    \phi_{0} = {0\choose\frac {v}{\sqrt{2}} }.
\end{equation}
With this choice we  break the symmetry. Notice that in the quantum case the constant $v$, having the dimension $M$,
is the vacuum expectation value (vev) of the Higgs field.

The doublet $\phi (x)$ can  be presented in the form
\begin{equation}
\begin{array}{l}
\phi(x) = e^{i\,\frac{1}{v}\frac{1}{2}\,\vec\tau\cdot\vec\theta (x)}~
\left(\begin{array}{c} 0\\\frac{v+H(x)}{\sqrt{2}}
\end{array}\right).
\end{array}
\label{Hpresent}
\end{equation}
Here  $\theta_{i} (x)$ ($i=1,2,3)$ and $H(x)$ are real functions which have dimension of the scalar field ($M$). Vacuum values of these functions are equal to zero.

The Lagrangian (\ref{HiggsL}) is invariant under the $SU_{L}(2)\times U_{Y}(1)$ local gauge transformations. We can choose the arbitrary gauge in such a way that
\begin{equation}
\begin{array}{l}
\phi(x) =
\left(\begin{array}{c} 0\\\frac{v+H(x)}{\sqrt{2}}
\end{array}\right).
\end{array}
\label{Higgsfield}
\end{equation}
Such a gauge is called the unitary gauge. From (\ref{Higgsfield}) for the Lagrangian (\ref{HiggsL}) we find the following expression
\begin{equation}\label{hL4}
\mathcal{L}=\frac{1}{2} ~\partial_{\alpha}H\,\partial^{\alpha}H
+\frac{1}{4}\, (v +H)^{2}g^{2} W^{\dag}_{\alpha}\,W^{\alpha}+
\frac{1}{4}\,(v +H)^{2}(g^{2}+g'^{2})\frac{1}{2}Z_{\alpha}\,Z^{\alpha}-
\frac{\lambda}{4}\, (2v H +H^{2})^{2}.
\end{equation}
The mass terms of $W^{\pm}$ and $Z^{0}$ vector bosons and  the scalar Higgs boson have the following form
\begin{equation}\label{massterm}
\mathcal{L}^{m}=m^{2}_{W}\,W^{\dag}_{\alpha}\,W^{\alpha} +
\frac{1}{2}\,m^{2}_{Z}\,Z_{\alpha}\,Z^{\alpha} -
\frac{1}{2}\,m^{2}_{H}\,H^{2},
\end{equation}
where $m_{W}$, $m_{Z}$ and $m_{H}$ are  masses of the $W^{\pm}$, $Z^{0}$ and  Higgs bosons.
From (\ref{hL4}) and (\ref{massterm}) we find
\begin{equation}\label{masses}
m_{W} =\frac{1}{2}\, g\, v ,\quad
m_{Z}=\frac{1}{2}\sqrt{(g^{2}+g^{'2})} v,\quad m_{H}=
\sqrt{2\lambda}v .
\end{equation}
Thus, after the spontaneous symmetry breaking  $W^{\alpha}(x)$, becomes
the field of the charged vector $W^{\pm}$ bosons with the mass
$\frac{1}{2}gv$, $Z^{\alpha}(x)$ becomes the field of neutral vector $Z^{0}$
bosons with the mass $\frac{1}{2}\sqrt{(g^{2}+g^{'2})
}~v$, $A_{\alpha}(x)$ remains the  field of massless photons.

 Three (Goldston) degrees of freedom are necessary to
provide longitudinal components of massive $W^{\pm}$ and $Z^{0}$ bosons. The Higgs doublet (two complex scalar fields, 4 degrees of freedom) is a {\em minimal possibility.} One remaining degree of freedom is a neutral  Higgs field $H(x)$
of scalar particles with the mass $\sqrt{2\lambda}~v$.

{\em The  Brout-Englert-Higgs mechanism of the generation of masses of  $W^{\pm}$ and $Z^{0}$ bosons  predicts an existence of the massive scalar boson.} Recent discovery of the scalar boson at LHC \cite{HiggsLHC} is an impressive confirmation of this prediction of the Standard Model.

The expressions (\ref{masses}) for masses of the $W^{\pm}$ and $Z^{0}$ bosons are characteristic expressions for masses of  vector bosons in a theory   with spontaneous symmetry breaking if in the  Lagrangian  covariant derivative of the Higgs field enters. In fact, it is evident from (\ref{HiggsL}) that masses of the vector bosons must have a form of a product of the constant part of the Higgs field (vacuum expectation value, $v$) and interaction constants.

The relations (\ref{masses}) allow to connect the constant $v$ with  the Fermi constant $G_{F}$. In fact, the Fermi constant, which can be determined from the measurement of time of life of muon and from other CC data, is given by the expression
\begin{equation}\label{Fermiconst}
    \frac{G_{F}}{\sqrt{2}}=\frac{g^{2}}{8m^{2}_{W}}.
\end{equation}
From (\ref{masses}) and (\ref{Fermiconst}) we obviously have
\begin{equation}\label{Fermiconst1}
v^{2}=\frac{1}{\sqrt{2}G_{F}}.
\end{equation}
Thus, we find
\begin{equation}\label{Fermiconst1}
v=(\sqrt{2}G_{F})^{-1/2}\simeq 246~\mathrm{GeV}.
\end{equation}
The interaction constant $g$ is connected with the electric charge $e$ and the parameter $\sin\theta_{W}$
by the unification condition (\ref{unif}). From (\ref{unif}), (\ref{masses}) and  (\ref{Fermiconst1}) for the mass of the $W$ boson we find the following expression
\begin{equation}\label{mwmass}
 m_{W}=(\frac{\pi\alpha}{\sqrt{2}G_{F}})^{1/2}\frac{1}{\sin\theta_{W}},
\end{equation}
where $\alpha\simeq \frac{1}{137.036}$ is the fine-structure constant. For the mass of the $Z^{0}$ boson we have
\begin{equation}\label{mzmass}
m_{Z}=\frac{m_{W}}{\cos\theta_{W}}=(\frac{\pi\alpha}{\sqrt{2}G_{F}})^{1/2}\frac{1}{\sin\theta_{W}\cos\theta_{W}}.
\end{equation}
The parameter $\sin^{2}\theta_{W}$ can be determined from the study of NC weak processes. From existing data  it was found the value $\sin^{2}\theta_{W}=0.23116(12)$  \cite{PDG}.

Thus, the Standard Model predicts masses of the $W^{\pm}$ and $Z^{0}$ bosons. These predictions are in a perfect agreement with experimental data. For average of all measured values of $m_{W}$ and $m_{Z}$ we have \cite{PDG}:
\begin{equation}\label{mW,Z}
 m_{W}=80.420 \pm 0.031 ~\mathrm{GeV},\quad  m_{Z}=91,1876\pm 0.0021  ~\mathrm{GeV}.
\end{equation}
 From the Standard Model (taking into account radiative corrections) we have
\begin{equation}\label{1mW,Z}
 m_{W}=80.381 \pm 0.014 ~\mathrm{GeV},\quad  m_{Z}=91,1874\pm 0.0021  ~\mathrm{GeV}.
\end{equation}
This agreement of the experimental data with one of the basic prediction of the SM is an important confirmation of the idea of the spontaneous breaking of the electroweak symmetry.

 We will consider now the Higgs mechanism of the generation of masses of the fundamental fermions. The fermion mass term is a Lorenz-invariant product of left-handed and right-handed components.
 It must be generated by a $SU_{L}(2)\times U_{Y}(1)$ invariant Lagrangian (after the spontaneous symmetry breaking). Let us  first consider the charged leptons. The most general Lagrangian which can generate the mass term of the charged leptons has the following Yukawa form
\begin{equation}\label{Lcharlep1}
\mathcal{L}^{lep}_{Y}=-\sqrt{2}\sum_{l_{1},l_{2}} \bar\psi^{lep}_{ l_{1} L}Y_{l_{1}l_{2}}l'_{2R}~\phi+\rm{h.c},
\end{equation}
where $Y$ is a $3\times3$ complex nondiagonal matrix. The Standard Model does not put any constraints on the matrix $Y$. The elements of this matrix are parameters of the SM.

After the spontaneous breaking of the symmetry from (\ref{lepSU2dub}),
(\ref{vev6}) and (\ref{Lcharlep1}) we have
\begin{equation}\label{Lcharlep2}
 \mathcal{L}^{lep}_{Y}=-\sum_{l_{1},l_{2}}\bar l'_{1L}Y_{l_{1}l_{2}}l'_{2R}(v+H)+\rm{h.c}.
\end{equation}
The proportional to $v$ term is the  mass term of charged leptons. In order to present it in the canonical form we need to diagonalize the matrix   $Y$. The general complex matrix $Y$ can be diagonalized by the biunitary transformation
\begin{equation}\label{Lcharlep3}
    Y=V_{L}~y~V^{\dag}_{R},
\end{equation}
where $V_{L}$ and $V_{R}$ are unitary matrices and $y$ is a diagonal matrix with positive diagonal elements.
From (\ref{Lcharlep2}) and (\ref{Lcharlep3}) we find
\begin{equation}\label{Lcharlep4}
\mathcal{L}^{lep}_{Y}=-\sum_{l=e,\mu,\tau}\bar l_{L}m_{l}l_{R}~(1+\frac{1}{v}H)+\rm{h.c}
=-\sum_{l=e,\mu,\tau} m_{l}\bar l~l~(1+\frac{1}{v}H).
\end{equation}
Here
\begin{equation}\label{Lcharlep5}
 l_{L}=\sum_{l_{1}}( V^{\dag}_{L})_{ll_{1}}~l'_{1L},\quad l_{R}=\sum_{l_{1}}( V^{\dag}_{R})_{ll_{1}}~l'_{1R},\quad
l=l_{L}+l_{R}
\end{equation}
and
\begin{equation}\label{Lcharlep6}
    m_{l}=y_{l}~v.
\end{equation}
The first term of the Lagrangian (\ref{Lcharlep4}) is the  mass term of the charged leptons. Thus, $l(x)$ is the field of the charged lepton $l$  with the mass $m_{l}$ ($l=e,\mu,\tau$). Left-handed and right-handed components of the fields of leptons  with definite masses are connected, correspondingly, with primed left-handed fields, components of the doublets $\psi^{lep}_{l L}(x)$, and right-handed fields, singlets, by the unitary transformations (\ref{Lcharlep5}).

The second term of  (\ref{Lcharlep4}) is the Lagrangian of the interaction of leptons and the Higgs boson
\begin{equation}\label{1Lcharlep}
\mathcal{L}^{lep}_{Y}=-\sum_{l=e,\mu,\tau}f_{l}~\bar l~l~H,
\end{equation}
where interaction constants are given by the relation
\begin{equation}\label{2Lcharlep}
f_{l}=\frac{1}{v}~m_{l}=(\sqrt{2}G_{F})^{1/2}m_{l}\simeq 4.06\cdot 10^{-3}
    \frac{m_{l}}{\mathrm{GeV}}.
\end{equation}
Let us express leptonic electromagnetic current and leptonic CC and NC through fields of leptons with definite masses $l(x)$. Taking into account the unitarity of the matrices $V_{L}$ and $V_{R}$ for the EM current we have
\begin{eqnarray}\label{1EM}
j^{EM}_{\alpha}&=&\sum_{l} (-1)\bar l'_{L}\gamma_{\alpha} l'_{L}+ \sum_{l} (-1)\bar l'_{R}\gamma_{\alpha} l'_{R}\nonumber\\&=&
\sum_{l} (-1)\bar l_{L}\gamma_{\alpha} l_{L}+ \sum_{l} (-1)\bar l_{R}\gamma_{\alpha} l_{R}=
\sum_{l} (-1)\bar l~\gamma_{\alpha}~ l
\end{eqnarray}
For the leptonic charged current we find
\begin{equation}\label{1CC}
j^{CC}_{\alpha}=2\sum_{l}\bar \nu'_{l'L}\gamma_{\alpha} l'_{L}=2\sum_{l}\bar \nu_{lL}\gamma_{\alpha} l_{L},
\end{equation}
where
\begin{equation}\label{2CC}
\nu_{lL}=\sum_{l_{1}}( V^{\dag}_{L})_{ll_{1}}~\nu'_{l_{1}L}.
\end{equation}
is the field of the flavor neutrino.
Finally, for the leptonic NC we obtain the following expression
\begin{eqnarray}
j^{NC}_{\alpha}&=&\sum_{l} \bar \nu'_{l'L}\gamma_{\alpha}  \nu'_{l'L}-\sum_{l}\bar l'_{L}\gamma_{\alpha} l'_{L}-2\sin^{2}\theta_{W}j^{EM}_{\alpha}  \\
   &=& \sum_{l} \bar \nu_{lL}\gamma_{\alpha}  \nu_{lL}-\sum_{l}\bar l_{L}\gamma_{\alpha} l_{L}-2\sin^{2}\theta_{W}j^{EM}_{\alpha}.
\end{eqnarray}
We will consider now briefly  the Brout-Englert-Higgs mechanism of the generation of masses of quarks. Let us assume that in the total Lagrangian enter the following $SU_{L}(2)\times U_{Y}(1)$ invariant Lagrangian of the Yukawa interaction of quark and Higgs fields
\begin{equation}\label{quarkH}
\mathcal{L}_{Y}^{\rm{quark}}=-\sqrt{2}\,\sum_{a,q_{1}=d,s,b}
\bar\psi_{aL}~Y_{a q_{1}}^{down}~ q'_{1R}~\phi -\sqrt{2}~\sum_{a,q_{1}=u,c,t
}\bar\psi _{aL}\, ~Y^{up}_{a q_{1}}\,q'_{1R} \,~ \tilde \phi+\rm{h.c.}
\end{equation}
Here
\begin{equation}\label{quarkH1}
\tilde{\phi}=i\,\tau_{2}\phi^{*}
\end{equation}
is the conjugated Higgs doublet and $Y_{a q_{1}}^{down}$ and  $Y_{a q_{1}}^{up}$ are $3\times 3$ complex nondiagonal matrices.

After the spontaneous breaking of the symmetry in the unitary gauge we have
\begin{equation}\label{quarkH2}
\phi(x) ={  0   \choose \frac{v+H(x)}{\sqrt{2}}} ,\quad
\tilde \phi(x) ={  \frac{v+H(x)}{\sqrt{2}}   \choose 0}.
\end{equation}
From (\ref{quarkH}) and (\ref{quarkH2}) we find
\begin{eqnarray}\label{quarkH4}
\mathcal{L}_{Y}^{\rm{quark}}&=&-\sum_{q_{1},q_{2}=d,s,b}
\bar q'_{1L}~Y_{q_{1} q'_{2}}^{down}~ q_{2R}~(v+H)\nonumber\\
&-&\sum_{q_{1},q_{2}=u,c,t}
\bar q'_{1L}~Y_{q_{1} q_{2}}^{up}~ q'_{2R} \,~ (v+H)+\rm{h.c.}
\end{eqnarray}
For the complex matrices $Y^{down}$ and $Y^{up}$ we have
\begin{equation}\label{quarkH5}
Y^{down}=V^{down}_{L}~y^{down}~V^{down\dag}_{R},\quad Y^{up}=V^{up}_{L}~y^{up}~V^{up\dag}_{R}.
\end{equation}
Here $V^{down}_{L,R}$ and $V^{up}_{L,R}$ are unitary matrices and $y^{down}$, $y^{up}$ are diagonal matrices with positive diagonal elements.

Using (\ref{quarkH5}) for the Lagrangian $\mathcal{L}_{Y}^{\rm{quark}}$ we find
\begin{equation}\label{quarkH6}
\mathcal{L}_{Y}^{\rm{quark}}=-\sum_{q=u,d,c,s,t,b} m_{q}~\bar q~q~(1+\frac{1}{v}H).
\end{equation}
Here
\begin{equation}\label{quarkH7}
    m_{q}=y_{q}~v,\quad q=u,d,c,s,t,b
\end{equation}
are masses of the quarks,
\begin{equation}\label{quarkH8}
    q_{L}=\sum_{q_{1}=d,s,b}(V_{L}^{down\dag})_{q q_{1}}~q'_{1L}~~(q=d,s,b)\quad  q_{L}=\sum_{q_{1}=u,c,t}(V_{L}^{up\dag})_{q q_{1}}~q'_{1L}~~(q=u,c,t)
\end{equation}
and
\begin{equation}\label{quarkH9}
    q_{R}=\sum_{q_{1}=d,s,b}(V_{R}^{down\dag})_{q q_{1}}~q'_{1R}~~(q=d,s,b)\quad  q_{R}=\sum_{q_{1}=u,c,t}(V_{R}^{up\dag})_{q q_{1}}~q'_{1R}~~(q=u,c,t)
\end{equation}
The first terms in the r.h.s. of Eq. (\ref{quarkH6}) is the mass term of the quark
\begin{equation}\label{quarkH10}
\mathcal{L}_{m}^{\rm{quark}}=-\sum_{q=u,d,...} m_{q}\bar q~q.
\end{equation}
The second term
\begin{equation}\label{quarkH11}
\mathcal{L}_{H}^{\rm{quark}}= -\sum_{q=u,d
,...} f_{q}~\bar q q ~H
\end{equation}
is the Lagrangian of the interaction of quarks and the scalar Higgs boson. The interaction constants $f_{q}$
 are given by the relation
\begin{equation}\label{quarkH12}
    f_{q}=\frac{m_{q}}{v}=m_{q}(\sqrt{2}G_{F})^{1/2}\simeq 4.06\cdot 10^{-3}
    \frac{m_{q}}{\mathrm{GeV}}.
\end{equation}
Let us express  the electromagnetic current, neutral current and charged current of quarks in terms of the fields of physical quarks with definite masses. Taking into account the unitarity of the matrices $V^{up}_{L,R}$ and $V^{down}_{L,R}$ for the electromagnetic current of quarks we have the following expression
\begin{equation}\label{qcurrents1}
j^{\mathrm{EM}} _{\alpha}= \frac{2}{3}\sum_{q=u,c,t}\bar q'\gamma_{\alpha} q'+(-\frac{1}{3})\sum_{q=d,s,b}\bar q'\gamma_{\alpha} q'=\sum_{q=u,d,...}e_{q}~\bar q\gamma_{\alpha} q,
\end{equation}
where $e_{q}=\frac{2}{3}~\mathrm{for}~ q=u,c,t$ and $e_{q}=-\frac{1}{3}~\mathrm{for}~ q=d,s,b$.

Analogously, for the the neutral current of quarks we find
\begin{eqnarray}\label{qcurrents2}
j^{\rm{NC}}_{\alpha}&=&\sum_{q=u,c,t}\bar q'_{L}\gamma_{\alpha} q'_{L}-\sum_{q=d,s,b}\bar q'_{L}\gamma_{\alpha} q'_{L}-2\sin^{2}\theta_{W}j^{\mathrm{EM}} _{\alpha}\nonumber\\
&=&\sum_{q=u,c,t}\bar q_{L}\gamma_{\alpha} q_{L}-\sum_{q=d,s,b}\bar q_{L}\gamma_{\alpha} q_{L}-2\sin^{2}\theta_{W}j^{\mathrm{EM}} _{\alpha}.
\end{eqnarray}
Thus, NC of the Standard Model is diagonal over quark fields (conserves quark flavor).

Finally, for the charged current of quarks we have
\begin{equation}\label{qcurrents3}
 j^{\rm{CC}}_{\alpha}=\bar u'_{L}\gamma_{\alpha} d'_{L}+\bar c'_{L}\gamma_{\alpha} s'_{L}+\bar t'_{L}\gamma_{\alpha} b'_{L}=\bar u_{L}\gamma_{\alpha} d^{\mathrm{mix}}_{L}+\bar c_{L}\gamma_{\alpha} s^{\mathrm{mix}}_{L}+
 \bar t_{L}\gamma_{\alpha} b^{\mathrm{mix}}_{L}.
\end{equation}
Here
\begin{equation}\label{qcurrents4}
d^{\mathrm{mix}}_{L}=\sum_{q=d,s,b}V_{uq}q_{L},~~s^{\mathrm{mix}}_{L}=\sum_{q=d,s,b}V_{cq}q_{L},~~b^{\mathrm{mix}}_{L}=
\sum_{q=d,s,b}V_{tq}q_{L}
\end{equation}
are "mixed fields". The matrix $V=V^{up}_{L}~V^{down\dag}_{L}$ is a unitary $3\times3$  Cabibbo-Kobayashi-Maskawa (CKM) matrix. Thus, the fields of down quarks enter into CC in the mixed form. The mixing is connected with the fact that the unitary matrices $V^{up}_{L}$ and $V^{down}_{L}$ are different.

The CKM matrix is characterized by three mixing angles $\theta_{12}$,
$\theta_{23}$, $\theta_{13}$ and one phase $\delta$ responsible for the $CP$ violation in the quark sector. It can be presented  in the following form
\begin{eqnarray}
V=\left(\begin{array}{ccc}c_{13}c_{12}&c_{13}s_{12}&s_{13}e^{-i\delta}\\
-c_{23}s_{12}-s_{23}c_{12}s_{13}e^{i\delta}&
c_{23}c_{12}-s_{23}s_{12}s_{13}e^{i\delta}&c_{13}s_{23}\\
s_{23}s_{12}-c_{23}c_{12}s_{13}e^{i\delta}&
-s_{23}c_{12}-c_{23}s_{12}s_{13}e^{i\delta}&c_{13}c_{23}
\end{array}\right)
\label{unitmixU1}
\end{eqnarray}
Here $c_{ik}=\cos\theta_{ik}$ and $s_{ik}=\sin\theta_{ik}$

From existing data all matrix elements of CKM matrix are known.
From the global fit of the data of numerous experiments  it was found \cite{PDG}
\begin{eqnarray}
|V|=\left(\begin{array}{ccc}0.97427\pm 0.00015 &0.22534\pm 0.00065&0.00351\pm 0.00015\\
0.22520\pm 0.00065&
0.97344\pm 0.00016&0.0412^{+0.0011}_{-0.0005}\\
0.00867^{+0.00029}_{-0.00031}&
0.0404^{+0.0011}_{-0.0005}&0.999146^{+0.000021}_{-0.000046}
\end{array}\right)
\label{unitmixU2}
\end{eqnarray}
From (\ref{2Lcharlep}) and (\ref{quarkH12}) for the masses of charged leptons and quarks we have
\begin{equation}\label{lqmasses}
m_{l}=f_{l}v,\quad m_{q}=f_{q}v.
\end{equation}
Thus, masses of leptons (quarks) have the form of the product of vev (coming from the Higgs field) and the constants of interaction of the leptons (quarks) and the Higgs boson. Let us notice that masses of $W^{\pm}$  and $Z^{0}$) vector bosons have the same form (see (\ref{masses}).

Masses of leptons and quarks are known. From (\ref{lqmasses}) follows that {\em  the SM predicts the constants of interaction of leptons and quarks with the Higgs boson.} The first LHC measurements of the constants  $f_{\tau}$ and  $f_{b}$ are in agreement with the SM prediction (see \cite{CMS}).

Up to now we considered the SM mechanism of the generation of masses of charged leptons and quarks. What about neutrinos? As we have discussed earlier, in the minimal Standard Model there are no right-handed neutrino fields. Thus, do not exist Yukawa interaction of neutrino fields with the Higgs doublet. This means that neutrino masses can not be generated by the Standard Higgs mechanism. {\em After spontaneous breaking of the electroweak symmetry neutrinos remain massless Weyl particles.}

This suggestion is supported by the comparison of masses of quarks, leptons and neutrinos. Let us consider particles of the third generation. We have
\begin{equation}\label{3generation}
m_{\tau}\simeq 1.77~\mathrm{GeV},\quad m_{t}\simeq 173~\mathrm{GeV},\quad m_{b}\simeq 4.2~\mathrm{GeV}
\end{equation}
Absolute values of neutrino masses are not known at present. From existing tritium Mainz \cite{Mainz} and Troitsk \cite{Troitsk} the following upper bounds were found,respectively,
\begin{equation}\label{mainz}
    m_{\beta}<2.3~\mathrm{eV},\quad m_{\beta}<2.05~\mathrm{eV},
\end{equation}
where $m_{\beta}=\sum_{i}\sqrt{|U_{ei}|^{2}m^{2}_{i}}$. If we assume the hierarchy of neutrino masses
($m_{1}\ll m_{2}\ll m_{3}$) from the neutrino oscillation data we have $m_{3}\simeq \sqrt{\Delta m^{2}_{A}}\simeq 5\cdot 10^{-2}$ eV ($\Delta m^{2}_{A}$ is the atmospheric neutrino mass-squared difference). Thus, masses of quarks and the lepton
of the third generation differs by  two orders of magnitude or less. The mass of neutrino $\nu_{3}$ is about 11 orders of magnitude smaller that the mass of the $\tau$-lepton.
{\em It is very unlikely that neutrino masses are of  the same SM origin as masses of quarks and charged leptons.}

We will present now   the arguments in favor of a beyond the SM origin of the neutrino masses in a different form. Let us assume that not only left-handed neutrino fields  $\nu'_{lL}$ but also right-handed  fields  $\nu'_{lR}$
are SM fields. The $SU_{L}(2)\times U_{Y}(1)$ invariant Lagrangian of the Yukawa interaction of neutrino and Higgs fields has the form
\begin{equation}\label{nuYukawa}
\mathcal{L}^{\nu}_{Y}=-\sqrt{2}\sum_{l'l}\overline  \psi^{lep}_{l_{1}L}Y^{\nu}_{l_{1}l_{2}}\nu'_{l_{2}R}\tilde{\phi }+\mathrm{h.c.}.
\end{equation}
After spontaneous breaking of the electroweak symmetry from (\ref{nuYukawa}) for the neutrino mass term
we find the following expression
\begin{equation}\label{Dmass}
 \mathcal{L}^{\mathrm{D}}=-v\sum_{l',l}\bar\nu'_{l'L}\,Y^{\nu}_{l'l} \nu'_{lR}
+\mathrm{h.c.}=-\sum_{l',l}\bar\nu_{l'L}~M_{l'l}^{D}\nu_{lR}+\mathrm{h.c.}
\end{equation}
Here $M^{D}=vV^{\dag}_{L}Y^{\nu}$ where the matrix $V_{L}$ connects fields $\nu'_{lL}$ and flavor neutrino fields  $\nu_{lL}$ (see (\ref{2CC})).
After the standard diagonalization of the matrix $V^{\dag}_{L}Y^{\nu}$  we find
 \begin{equation}\label{Dmass1}
\mathcal{L}^{\mathrm{D}}=\sum^{3}_{i=1}m_{i}\bar \nu_{i}\nu_{i} ,\quad \nu_{lL}=\sum_{i} U_{li}\nu_{iL},
 \end{equation}
where $U$ is a unitary mixing matrix and $\nu_{i}$ is a field of the Dirac neutrinos with mass $m_{i}$. For
neutrino mass we have
\begin{equation}\label{Dmass2}
    m_{i}=v~y_{i},
\end{equation}
where $y_{i}$ is the Yukawa coupling.

Let us consider the third, heaviest family. Assuming hierarchy of neutrino masses from neutrino oscillation data we have $m_{3}\simeq 5\cdot 10^{-2}$  eV. Thus, from (\ref{Dmass2}) we find $y_{3}\simeq 10^{-13}$. Yukawa couplings of quarks and lepton of the  third generation are in the range
( $1- 10^{-2}$). Extremely small values of the neutrino Yukawa couplings is commonly considered as a strong argument against SM origin of the neutrino masses.

\section{Majorana neutrino masses}
A mass term is a Lorenz-invariant product of the left-handed and right-handed components of the fields. Is it possible to have neutrino mass term in the case if there are no right-handed  fields? For the simplest case of two neutrinos this problem was solved many years ago in the paper \cite{GribPonte}. The only possibility to built the neutrino mass term in this case is to assume that the total lepton number $L$ is not conserved. In fact, it is easy to see that the conjugated field $(\nu_{lL})^{c}=
C(\bar\nu_{lL})^{T}$ ($C$ is the matrix of the charge conjugation which satisfies the conditions $C\gamma^{T}_{\alpha}C^{-1}=-\gamma_{\alpha},~ C^{T}=-C$) is the right-handed field.\footnote{Taking into account that $C\gamma^{T}_{5}C^{-1}=\gamma_{5}$ from the definition of left-handed (right-handed) field ($\gamma_{5}~\psi_{L,R}=\mp~\psi_{L,R}$) we have $\gamma_{5} (C~\bar\psi_{L,R}^{T})=\pm~(C\bar\psi_{L,R}^{T})$.}
Thus, assuming that the total lepton number is not conserved in the case of the left-handed neutrino fields $\nu_{lL}$ for the (Majorana) neutrino mass term we have
\begin{equation}\label{Mjmass}
\mathcal{L}^{M}=-\frac{1}{2}\sum_{l_{1},l_{2}}
 \bar\nu_{l_{1}L}~ M_{l_{1}l_{2}}^{M}~ (\nu_{l_{2}L})^{c}+\mathrm{h.c.},
\end{equation}
where  $M^{M}$ is a $3\times 3$ complex matrix. From general requirement of the Fermi-Dirac statistics it follows that
$M^{M}=(M^{M})^{T}$.\footnote{In fact, we have $$\sum_{l_{1},l_{2}}
\bar\nu_{l_{1}L}~ M^{M}_{l_{1}l_{2}}~ (\nu_{l_{2}L})^{c}=-\sum_{l_{1},l_{2}}
 \bar\nu_{l_{2}L}~ M^{M}_{l_{1}l_{2}}~ C^{T}\bar\nu_{l_{2}L}^{T}=\sum_{l_{1},l_{2}}
 \bar\nu_{l_{1}L}~ (M^{M})^{T}_{l_{1}l_{2}})^{T}~ (\nu_{l_{2}L})^{c}.$$}
A complex symmetrical matrix $M^{M}$ can be diagonalized with the help of one unitary matrix. We have
\begin{equation}\label{diagon}
     M^{M}=U~ m~ U^{\dag},
\end{equation}
where $U^{\dag}~U=1$, $m_{ik}=m_{i}\delta_{ik},~~m_{i}>0$.

From (\ref{Mjmass}) and (\ref{diagon}) we obtain the  Majorana mass term in the diagonal form
\begin{equation}\label{Mjmass2}
\mathcal{L}^{M}=-\frac{1}{2}\sum^{3}_{i=1}m_{i}~\bar \nu_{i}\nu_{i},
\end{equation}
where
\begin{equation}\label{Mjcond}
\nu_{i}=\nu^{c}_{i}=C\bar \nu_{i}^{T}.
\end{equation}
The flavor field  $\nu_{lL}$ is connected with the fields of Majorana neutrinos with definite masses $\nu_{i}$ by the standard mixing relation
\begin{equation}\label{mix}
 \nu_{lL} =\sum^{3}_{i=1}U_{li}~\nu_{iL},
\end{equation}
where $U$ is the unitary PMNS mixing matrix.

The Majorana mass term (\ref{Mjmass}) can be generated only in the framework of a beyond the Standard Model physics.
The method of the effective Lagrangians \cite{Weinberg,Wilczek} is the most general way
which allows to describe effects  of a beyond the SM physics.\footnote{The classical example of the effective Lagrangian is the four-fermion Fermi Lagrangian of the $\beta$-decay.} The effective Lagrangian is a nonrenormalizable  Lagrangian
which is built from SM fields,  have dimension five or more  and is invariant under electroweak $SU_{L}(2)\times U_{Y}(1)$ transformations.

In order to generate the neutrino mass term we need to build the effective Lagrangian which is quadratic in the lepton fields. The term
$(\bar \psi^{lep}_{lL}\tilde{\phi })$ ($l=e,\mu,\tau$) is $SU_{L}(2)\times U_{Y}(1)$ invariant and has dimension $M^{5/2}$. After spontaneous breaking of the symmetry it contains the  left-handed neutrino field multiplied by vev.
The lepton number violating  $SU_{L}(2)\times U_{Y}(1)$  invariant Lagrangian quadratic in the lepton fields has the form \cite{Weinberg}
\begin{equation}\label{effective}
 \mathcal{L}_{I}^{\mathrm{eff}}=-\frac{1}{\Lambda}~\sum_{l_{1},l_{2}}(\bar \psi^{lep}_{l_{1}L}\tilde{\phi })~ Y'_{l_{1}l_{2}}~(\tilde{\phi }^{T} (\psi^{lep}_{l_{2}L})^{c})+\mathrm{h.c.}.
\end{equation}
Here $\Lambda$ is a parameter which has the dimension $M$ and $Y'=(Y')^{T}$ is a dimesionless $3\times3$ matrix.

The the parameter $\Lambda$ characterizes the scale a new physics  at which the lepton number $L$ is violated. It is natural to expect that  $\Lambda\gg v$.

After the spontaneous breaking of the electroweak symmetry from from (\ref{quarkH2}) and (\ref{effective}) we have
\begin{equation}\label{effective1}
\mathcal{L}_{I}^{\mathrm{eff}}=-\frac{1}{2\Lambda} \sum_{l_{1},l_{2}}
 \bar\nu'_{l_{1}L}~ Y'_{l_{1}l_{2}}~ (\nu'_{l_{2}L})^{c}~(v+H)^{2}+\mathrm{h.c.}
\end{equation}
The term proportional to $v^{2}$ is the Majorana mass term. The flavor neutrino fields $\nu_{lL}$, which enter into the leptonic CC, are determined by the relation
 (\ref{2CC}). In terms of the flavor neutrino fields from (\ref{effective1}) we obtain the Majorana mass term (\ref{Mjmass}) in which the matrix  $M^{M}$ is given by
\begin{equation}\label{Mjmat}
M^{M}=\frac{v^{2}}{\Lambda}~\bar Y,
\end{equation}
where
\begin{equation}\label{Mjmat1}
\bar Y=V^{\dag}_{L}Y'(V^{\dag}_{L})^{T}
\end{equation}
is a symmetrical matrix.

From (\ref{Mjmat}) for the mass of the Majorana neutrino $\nu_{i}$ we find the following expression
\begin{equation}\label{Mjmat2}
m_{i}=\frac{v}{\Lambda}~\bar y_{i}v,\quad i=1,2,3,
\end{equation}
where $\bar y_{i}$ is the eigenvalue of the matrix $\bar Y$.

Thus, Majorana neutrino mass $m_{i}$ generated by the effective Lagrangian (\ref{effective}) is
a product of a "typical fermion mass" $v~\bar y_{i}$ and a suppression factor which is given by the ratio of the electroweak scale $v$ and a scale $\Lambda$ of a new lepton-number violating physics. The scheme based on the effective Lagrangian approach is a natural framework for the generation of neutrino masses which are much smaller than the masses of other fundamental fermions. Let us stress that such a scheme does not put any constraints of the mixing matrix $U$.

Absolute value of neutrino masses are not known at present. If we assume hierarchy of neutrino masses  for the mass of the heaviest neutrino  we have $m_{3}\simeq \sqrt{\Delta m^{2}_{A}}\simeq 5\cdot 10^{-2}$ eV. Assuming also that $\bar y_{3}\simeq 1$ for a new scale $\Lambda$ we find the estimate $\Lambda\simeq 10^{15}$ GeV. Thus, small Majorana neutrino masses could be a signature of a new very large lepton number violating scale in physics.

Let us summarize our discussion of the generation of the neutrino masses  based on the effective Lagrangian approach.
 \begin{enumerate}
   \item There is one possible lepton number violating effective Lagrangian which (after spontaneous breaking of the symmetry) leads to the Majorana neutrino mass term.
   \item  Number of Majorana neutrinos with definite masses is equal to the number of the flavor neutrinos (three).

\item Smallness of the neutrino masses with respect to the masses of quarks and leptons could signify  existence of a new lepton number violating scale  which is much larger than the electroweak scale $v$.

 \item  The Lagrangian (\ref{effective}) is the only possible effective Lagrangian of the dimension 5 (proportional to $\frac{1}{\Lambda}$). Other effective Lagrangians have dimension 6 and higher and at the very large $\Lambda$ give much smaller contributions to observables. Thus, neutrino masses are the most sensitive probe of a new physics at a scale which is much larger than the electroweak scale.
 \end{enumerate}
 Let us notice that or the dimensional arguments we used  it is important that Higgs is not composite particle and {\em exist Higgs field having dimension $M$}. Recent discovery of the Higgs boson at CERN \cite{Higgs} confirm this assumption.

 The violation of the lepton number can be connected with the existence of heavy Majorana leptons which interact with the lepton -Higgs pairs \cite{seesaw,Weinberg80}. Let us assume that heavy Majorana leptons $N_{i}$ with the mass $M_{i}\gg v$ ($N_{i}=N^{c}_{i},\quad i=1,...n$), singlets of $SU_{L}(2)\times U_{Y}(1)$ group, have the following Yukawa lepton-number violating interaction
\begin{equation}\label{HeavyMj}
 \mathcal{L}_{I}^{Y}=-\sqrt{2}\sum _{l,i}y_{li} \bar \psi^{lep}_{lL}\tilde{\phi }N_{iR}+\mathrm{h.c.}.
\end{equation}
Here  $y_{li}$ are dimensionless Yukawa coupling constants.

In the second order of the perturbation theory at the electroweak energies  the interaction (\ref{HeavyMj}) generates the effective Lagrangian (\ref{effective}) in which the constant $\frac{\bar Y_{l_{1}l_{2}}}{\Lambda}$ is given by the relation
\begin{equation}\label{HeavyMj1}
 \frac{\bar Y_{l_{1}l_{2}}}{\Lambda}=\sum_{i}y_{l_{1}i}\frac{1}{M_{i}} y_{l_{2
 }i}.
\end{equation}
Thus, the scale of a new lepton number violating physics is determined by masses of the heavy Majorana leptons.

We have considered  a beyond the SM mechanism of the generation of neutrino masses and mixing based on idea of violation of the lepton number at a  scale which is much larger that the electroweak scale. This approach does not allow to predict values of the neutrino masses, neutrino mixing angles and $CP$-violating phases. However, there are two general
consequences of this mechanism.

\begin{enumerate}

\item {\em Neutrino with definite masses $\nu_{i}$ must be Majorana particles.} Nature of neutrinos with definite masses is a fundamental unsolved experimental problem. As it is well known, the most practical way to reveal the nature of neutrinos with definite masses (Majorana or Dirac?) is to look for neutrinoless double $\beta$-decay ($0\nu\beta\beta$-decay) of some even-even nuclei
\begin{equation}\label{betabeta}
(A,Z)\to (A,Z+2)+ e^{-}+e^{-}.
\end{equation}
In the case of three Majorana neutrino mixing the total decay-rate of the $0\nu\beta\beta$-decay have the following general form
\begin{equation}\label{totrate}
\Gamma^{0\nu}=\frac{1}{T^{0\nu}_{1/2}}(Z)=|m_{\beta\beta}|^{2}~|M^{0\nu}(Z)|^{2}~
G^{0\nu}(Q,Z).
\end{equation}
Here
\begin{equation}\label{effMaj}
m_{\beta\beta}=\sum_{i}U^{2}_{ei}m_{i}
\end{equation}
is the effective Majorana mass, $M^{0\nu}(Z)$ is the nuclear matrix element (NME), which is determined by the nuclear properties and does not depend on elements of the neutrino mixing matrix and small neutrino masses, and $G^{0\nu}(Q,Z)$ is known phase space factor which includes the Fermi function describing final state Coulomb interaction of two electrons and the nuclei. The calculation of NME is a very complicated nuclear problem. NME for the $0\nu\beta\beta$-decay of different nuclei were calculated in the framework of several many-body approximate schemes.
Results of these calculations  for some nuclei differ in 2-3 times. A  progress in improving of the existing calculations of NME is expecting (see \cite{Engel}).

Several experiments on the search for the $0\nu\beta\beta$-decay of different nuclei are going on  at present. Up to now only lower bounds on half-life of the $0\nu\beta\beta$-decay were obtained. Recently several new results were reported.

In the  EXO-200 experiment \cite{Exo} the $0\nu\beta\beta$-decay of $^{136}\mathrm{Xe}$ ($Q=2458$ KeV) was investigated in the liquid  time-projection chamber (with 80.6\% enrichment in $^{136}\mathrm{Xe}$). After $100~ kg\cdot y$ exposure the following lower bound was obtained
\begin{equation}\label{Exo}
    T^{0\nu}_{1/2}(^{136}\mathrm{Xe})>1.1\cdot 10^{25}~\mathrm{y}\quad (90\% CL)
\end{equation}
Using different calculations of NME from this result for the effective Majorana mass the following upper bounds were found
\begin{equation}\label{Exo1}
|m_{\beta\beta}|<(1.9-4.5)\cdot 10^{-1}~\mathrm{eV}
\end{equation}
In the KamLAND-Zen experiment \cite{KZen} 383 kg of liquid $^{136}\mathrm{Xe}$ (enriched to 90.77\% ) was loaded in the liquid scintillator. The $0\nu\beta\beta$-decay of $^{136}\mathrm{Xe}$ was searched for. After 115 days of exposure for the half-life the following lower bound was inferred
\begin{equation}\label{Kzen}
    T^{0\nu}_{1/2}(^{136}\mathrm{Xe})>1.3\cdot 10^{25}~\mathrm{y}\quad (90\% CL)
\end{equation}
Combining this result with the result of the previous run  for the half-life it was obtained
\begin{equation}\label{Kzen1}
    T^{0\nu}_{1/2}(^{136}\mathrm{Xe})>2.6\cdot 10^{25}~\mathrm{y}\quad (90\% CL)
\end{equation}
From this bound for the effective Majorana mass was found
\begin{equation}\label{Exo1}
|m_{\beta\beta}|<(1.4-2.8)\cdot 10^{-1}~\mathrm{eV}.
\end{equation}
In the germanium GERDA experiment \cite{Gerda} the $0\nu\beta\beta$-decay of $^{76}\mathrm{Ge}$ ($Q=2039$ KeV)
was studied. In the Phase-I of the experiment the germanium target mass was 21.6 kg (86\% enriched in $^{76}\mathrm{Ge}$). Very law background ($ 10^{-2}~cts/KeV~kg~y$) was reached. For the
the lower bound for the half-life of $^{76}\mathrm{Ge}$ it was obtained the value
\begin{equation}\label{Gerda}
    T^{0\nu}_{1/2}(^{76}\mathrm{Ge})>2.1\cdot 10^{25}~\mathrm{y}\quad (90\% CL).
\end{equation}
This result allowed the collaboration to refute the claim of the observation of the
$0\nu\beta\beta$-decay of $^{76}\mathrm{Ge}$ made in \cite{Klapdor}. Combining (\ref{Gerda}) with the results of Heidelberg-Moscow\cite{HeidMos} and IGEX \cite{Igex} experiments it was found
\begin{equation}\label{Gerda1}
    T^{0\nu}_{1/2}(^{76}\mathrm{Ge})>3.0\cdot 10^{25}~\mathrm{y}\quad (90\% CL).
\end{equation}
From this bound for the effective Majorana mass it was obtained the following bound
\begin{equation}\label{Gerda3}
|m_{\beta\beta}|<(2-4)\cdot 10^{-1}~\mathrm{eV}.
\end{equation}
Let us notice that in the next generation of the experiments on the search for the $0\nu\beta\beta$-decay
it is planned to reach the region
\begin{equation}\label{Gerda3}
|m_{\beta\beta}|<\mathrm{a~few}\cdot 10^{-2}~\mathrm{eV}.
\end{equation}

\item {\em The number of neutrinos with definite masses must be equal to the number of lepton-quark generations (three).} This means that there are no sterile neutrinos in such a scheme. At present exist some indications in favor of a fourth
neutrino mass-squared difference $\sim 1~\mathrm{eV}^{2}$ (see \cite{Giunti,Schwetz}). In the simplest 3+1 scheme (three light neutrinos and one neutrino with mass about 1 eV) for the short baseline experiments sensitive to $\Delta m^{2}_{14}$ we find the following expression for $\nua{\alpha}\to \nua{\alpha'}$ transition probability
\begin{equation}\label{transition1}
P(\nua{\alpha}\to \nua{\alpha'})=\delta_{\alpha,\alpha'}-4(\delta_{\alpha,\alpha'}-|U_{\alpha' 4}|^{2})|U_{\alpha' 4}|^{2}\sin^{2}{\frac{\Delta m^{2}_{14}L}{4E}},
\end{equation}
where $E$ is neutrino energy, $L$ is the source-detector distance, $\Delta m^{2}_{14}=m^{2}_{4}-m^{2}_{1}$. Positive indications in favor of  neutrino oscillations were found in the LSND $\nua{\mu}\to \nua{e}$
experiment \cite{LSND}, in short baseline reactor  $\bar\nu_{e}\to\bar\nu_{e}$ experiments (see \cite{reactor}) and in radioactive source $\nu_{e}\to\nu_{e}$ experiments \cite{sourse}. From (\ref{transition1}) it follows that oscillation amplitudes for $\nua{\mu}\to \nua{e}$  $\nua{e}\to \nua{e}$ and $\nu_{\mu}\to \nu_{\mu}$ transitions are given by the expressions
$A_{e\mu}=4|U_{e 4}|^{2}|U_{\mu 4}|^{2}$, $A_{e e}=4|U_{e 4}|^{2}(1-|U_{e 4}|^{2}\simeq 4|U_{e 4}|^{2}$ and
 $A_{\mu \mu}=4|U_{\mu 4}|^{2}(1-|U_{\mu 4}|^{2}\simeq 4|U_{\mu 4}|^{2}$. These relations mean that from analysis of $\nua{\mu}\to \nua{e}$  and $\nua{e}\to \nua{e}$ data  it is possible to obtain the allowed region for $\nu_{\mu}\to \nu_{\mu}$ oscillations. There are no indications in favor of $\nu_{\mu}\to \nu_{\mu}$  disappearance in
short baseline experiments \cite{CDHS,Minos}. Global analysis of the data of all short baseline experiments indicates a strong tension between data \cite{Schwetz}. The same problem exist if data are analyzed in the framework of more complicated models.

Several new experiments specially designed  to check existing
controversial indications in favor  for short baseline neutrino oscillations  are in preparation at present (see \cite{Rubbia}).

\end{enumerate}

If heavy Majorana leptons  exist their CP-violating decays  in the early Universe could be the origin of the lepton asymmetry which due to sphaleron effects can be transferred into the barion asymmetry of the Universe  \cite{leptogenesis}.

Let us notice that the mechanism based on the interaction (\ref{HeavyMj}) is called type I seesaw. The effective Lagrangian (\ref{effective}) can also be generated by the Lagrangian of interaction of lepton-Higgs doublets with a heavy triplet lepton (type III seesaw) and by the Lagrangian of interaction of lepton doublets and Higgs doublets with heavy triplet scalar boson (type II seesaw).

\section{Conclusion}
The Standard Model  successfully describes all observed physical phenomena in a wide range of energies up to a few hundreds GeV. After the discovery of the Higgs boson at LHC the Standard Model was established as  a correct theory of physical phenomena in the electroweak scale.

It is suggested here that neutrinos play exceptional role in the Standard Model. Neutrinos possibly are crucial in
the determination of symmetry properties of the SM. It is very plausible that neutrinos are  the only Standard Model particles whose properties (masses and  nature) are determined by a beyond the Standard Model physics.

The Standard Model is based on
\begin{itemize}
  \item The local gauge symmetry.
  \item The unification of the weak and electromagnetic interactions.
  \item Brout-Englert-Higgs mechanism of the spontaneous breaking of the symmetry.
\end{itemize}
It looks that in the framework of these general principles nature choose the simplest possibilities. The simplest, most economical possibility for neutrinos is to be massless two-component  Weyl particles (which is the Landau-Lee-Yang-Salam two-component neutrino). The experiment showed that from two possibilities (left-handed or right-handed) nature choose the left-handed possibility.

In order to ensure symmetry, fields of quarks and leptons also must be massless.  The symmetry group must be non-Abelian. This allow to include charged particles and ensure the universality of the minimal interaction of the fundamental fermions and the gauge fields. The simplest possibility is $SU_{L}(2)$ with doublets of the left-handed fields.

The unification of weak and electromagnetic interactions require enlargement of the symmetry group. The simplest possibility is $SU_{L}(2)\times U_{Y}(1)$ group. Because the electromagnetic current is the sum of the left-handed and right-handed parts, right-handed fields of the {\em charged particles} also must be  SM fields (singlets of the $SU_{L}(2)$ group). Electric charges of neutrinos are equal to zero. The unification principle does not requires existence of the right-handed neutrino fields. Minimal possibility (the basic principle of the SM): {\em there are no right-handed neutrino fields in the SM.} This is a crucial feature of the SM. Nonconservation of $P$ and $C$ in the weak interaction is connected with that. Because of there are no right-handed SM neutrino fields neutrinos are the only particles which after spontaneous breaking of the electroweak symmetry  remain massless.

Neutrinos can have only beyond the Standard Model Majorana masses (lepton number violating Majorana mass term).
{\em This is the most economical possibility.} It is generated by the unique, beyond the Standard Model dimension five Weinberg effective Lagrangian. Due to a suppression factor which is a ratio of the electroweak vev and a scale of a new lepton number violating physics such approach naturally explains the smallness of neutrino masses.

In the framework of the effective Lagrangian it is impossible to predict neutrino masses, mixing angles and phases.
The same is true for leptons and quarks: the Higgs mechanism of the generation of masses and mixing of leptons and quarks do not predicts the values of masses, mixing angles and phases.
However, there are three general consequences of the considered mechanism of the neutrino mass generation.
\begin{enumerate}
  \item Neutrino with definite masses $\nu_{i}$ are Majorana particles.
  \item Number of neutrinos with definite masses is equal to the number of the flavor neutrinos (three).
  \item The total lepton number $L$ is violated at a large scale $\Lambda$.
\end{enumerate}
The neutrino nature (Majorana or Dirac ?) can be inferred from the experiments
on the the search for neutrinoless double $\beta$-decay of $^{76} \mathrm{Ge}$, $^{136} \mathrm{Xe}$ and other nuclei. If this process will be observed it will be a proof that neutrinos with definite masses are Majorana particles, i.e. that neutrino masses are of a beyond the SM origin. Future experiments will probe inverted neutrino mass spectrum region ($m_{\beta\beta}\simeq \mathrm{a~few}10^{-2}$ eV).
In the case of normal mass hierarchy the probability of the neutrinoless double $\beta$-decay will be so small that new methods of the detection of the process must be developed (see \cite{futurebb}).

A possibility that the number of the neutrinos with definite masses is more than three will be tested in future
reactor, radioactive source and accelerator experiments on the search for sterile neutrinos.

If Yukawa constants are of the order of one the scale  $\Lambda$ is very large ($\simeq 10^{15}$~GeV). In this case a successful theory of  baryogenesis could be an indirect test of the point 3. Let us notice, however, that much smaller values of  $\Lambda$ ($\simeq$ TeV) were considered in numerous papers (see, for example, \cite{Conca}).

The Standard Model teaches us that the simplest possibilities are more likely to be correct. Massless two-component left-handed Weyl neutrinos and absence of the right-handed neutrino fields in the Standard Model is the simplest, most elegant and most economical possibility. In this case  beyond the Standard Model Majorana mass term  is a unique possibility for neutrinos to be massive and mixed.

This work is supported by the Alexander von Humboldt Stiftung, Bonn, Germany (contract Nr. 3.3-3-RUS/1002388),
by RFBR Grant N 13-02-01442 and by the Physics Department E15 of the Technical University Munich. I am thankful to W. Potzel for useful discussions and to the theory group of TRIUMF for the hospitality.

\end{document}